\documentclass[journal]{IEEEtran}
\usepackage{cite}

  \usepackage[caption=false,font=footnotesize]{subfig}
\usepackage{amsmath} % assumes amsmath package installed
\usepackage{amssymb}  % assumes amsmath package installed
\usepackage{amsthm}
\usepackage{color}
\usepackage{subfig}
\usepackage{graphicx}

\DeclareSymbolFont{bbold}{U}{bbold}{m}{n}
\DeclareSymbolFontAlphabet{\mathbbold}{bbold}
\usepackage{calrsfs}
\DeclareMathAlphabet{\pazocal}{OMS}{zplm}{m}{n}
\renewcommand{\mathcal}[1]{\pazocal{#1}}
\DeclareMathOperator*{\SIS}{SIS}

\usepackage{accents}
\newcommand{\ubar}[1]{\underaccent{\bar}{#1}} %keep in case the above does not 

\theoremstyle{plain}
\newtheorem{theorem}{Theorem}

\theoremstyle{definition}
\newtheorem{definition}{Definition}

\theoremstyle{remark}
\newtheorem{remark}{Remark}
\usepackage{mathtools}

\newcounter{MYtempeqncnt}

\allowdisplaybreaks

\ifdefined \twocolumnversion
\else
\usepackage{changepage}
\fi

\begin{document}
\title{Synergistic Effects in Networked Epidemic Spreading Dynamics}

\author{Masaki Ogura,~\IEEEmembership{Member,~IEEE,}
        Wenjie~Mei,
        and~Kenji~Sugimoto,~\IEEEmembership{Member,~IEEE}% <-this % stops a space
\thanks{M.~Ogura, W.~Mei, and K.~Sugimoto are with the Division of Information Science, Nara Institute of Science and Technology, Ikoma, Nara 630-0192, Japan. e-mail: \{oguram,\,mei.wenjie.mu2,\,kenji\}@is.naist.jp.}% <-this % stops a space
\thanks{This work was supported by JSPS KAKENHI Grant Number JP18K13777.}% <-this % stops a space
}


\markboth{Journal of \LaTeX\ Class Files,~Vol.~14, No.~8, August~2015}%
{Shell \MakeLowercase{\textit{et al.}}: Bare Demo of IEEEtran.cls for IEEE Journals}

\maketitle

\begin{abstract}
In this brief, we study epidemic spreading dynamics taking place in complex networks. We specifically investigate the effect of synergy, where multiple interactions between nodes result in a combined effect larger than the simple sum of their separate effects. Although synergistic effects play key roles in various {biological and social} phenomena, their analyses have been often performed by means of approximation techniques and for limited types of networks. {In order to address this limitation, this paper proposes a rigorous approach to quantitatively understand the effect of synergy in the Susceptible-Infected-Susceptible model taking place in an arbitrary complex network.} We derive an upper bound on the growth rate of the synergistic Susceptible-Infected-Susceptible model in terms of the eigenvalues of a matrix whose size grows quadratically with the number of the nodes in the network. {We confirm the effectiveness of our result by numerical simulations on empirically observed human and animal social networks.}
\end{abstract}


% \fi
\IEEEpeerreviewmaketitle

\section{Introduction}

\IEEEPARstart{U}{nderstanding} epidemic spreading dynamics taking place in complex networks is of utmost importance with applications in epidemiology \cite{Pastor-Satorras2015a}, opinion formation, rumor propagation, and cyber security~\cite{Xia2008,Babaei2011}. During the last two decades, several important advances have been made toward the analysis, modeling, and control of networked spreading processes~\cite{Nowzari2015a}. For example, fundamental connections between the eigenvalues of a network and epidemic thresholds were identified in~\cite{Ganesh2005}. A unified framework for the analysis of spreading processes evolving on multilayer networks was presented in~\cite{DarabiSahneh2013}. The dynamical change of network structures dependent on the evolution of spreading processes has been actively investigated in the context of adaptive networks~\cite{Zhu2018}. Recently, control and optimization tools have been proven to be effective for containing spreading processes in static~\cite{Preciado2014}, multi-layer~\cite{Zhao2018}, and temporal~\cite{Ogura2015c} networks. A prominent example of the achievements of network epidemiology can be found in, for example, effective predictions of the pandemic of H1N1 influenza on 2009~\cite{Balcan2009,Merler2011}.

All the aforementioned results assume nearest-neighbor interactions, in which individual nodes are directly influenced by only their neighbors. However, in realistic epidemic spreading processes, we frequently observe \emph{multi-node interactions}, in which a node can affect not only its neighbors but also other nodes such as the neighbors of its neighbors. A typical example of such phenomena can be found in the colonization by fungal and bacterial pathogens~\cite{Ben-Jacob1994} and tumor growth~\cite{Liotta2001}. In the context of human social networks, it has been experimentally confirmed that the spreads of opinions~\cite{Castellano2009} and behaviors~\cite{Centola2010} are effectively accelerated by \emph{synergistic effects}, {which make the combined effect from multiple interactions greater than the simple sum of the effect of individual interactions}.

We find in the literature several works providing theoretical insight for understanding the effect of synergistic interactions on spreading processes. The authors in~\cite{Perez-Reche2011} considered an extended Susceptible-Infected-Recovered (SIR) model with synergistic effects, and numerically confirmed on lattices that synergy can significantly alter asymptotic behaviors of the spread. Similar models were further investigated analytically on lattices~\cite{Taraskin2013} and small-world networks~\cite{Broder-Rodgers2015}, respectively. The authors in~\cite{Liu2017b} studied the effect of synergy in the Susceptible-Infected-Susceptible (SIS) model under the assumption that the nodes having the same degree behave in the same manner. The effect of synergy on deterministic spreading processes was investigated in~\cite{Juul2017} with a mean-field approximation technique.

This brief proposes a rigorous but tractable approach to quantitatively analyze the effect of synergy in the SIS model taking place in \emph{arbitrary} networks, \emph{without} resorting to mean-field approximations. A primary challenge in this context is to obtain computationally tractable results, because the state space of the spreading model grows exponentially with the number of nodes in the network. To overcome this difficulty, in this brief we show that the growth rate of the size of the infected population in the network can be bounded from above by the maximum real part of the eigenvalues of a matrix whose size grows only \emph{quadratically} with the number of the nodes. In the derivation of the upper bound, we use Poisson-type stochastic differential equations describing the spreading model. The effectiveness of our theoretical results is numerically illustrated by simulations on empirical human and animal social networks.

This brief is organized as follows. After preparing mathematical notations, in Section~\ref{sec:SSIS} we introduce the synergistic SIS model over arbitrary complex networks. In Section~\ref{sec:grwoth}, we derive our upper-bound on the growth rate of the model. Numerical simulations are presented in Section~\ref{sec:sim}.

\subsubsection*{Notations} 
The maximum real part of the eigenvalues of a square matrix~$A$ is denoted by $\lambda_{\max}(A)$. An undirected graph is defined as the pair~$G=(V,E)$, where $V=\{ {1},\dotsc,{N}\} $ is a set of nodes and $E$ is a set of edges, i.e., unordered pairs of nodes. We say that nodes~$i$ and~$j$ are adjacent if $\{i, j\}\in\mathcal E$. The neighborhood of node~$i$, denoted by $\mathcal N_i$, is defined as the set of nodes adjacent to node~$i$. The adjacency matrix~$A=[a_{ij}]_{i, j}$ of~$G$ is defined as the $N\times N$ matrix such that $a_{ij} = 1$ if nodes~$i$ and~$j$ are adjacent, and $a_{ij}=0$ otherwise.

\section{Synergistic SIS model}\label{sec:SSIS}

We introduce the synergistic SIS model over complex networks in this section. Let us start by reviewing the standard SIS model without synergy (see, e.g., \cite{VanMieghem2009a}). Let $G$ be an undirected graph having $N$ nodes {labeled as $i = 1, \dotsc, N$}, with edges representing interactions between nodes. At time~$t \geq 0$, each node can be either \emph{susceptible} or \emph{infected}. When a node~$i$ is infected, the node transitions into the susceptible state with an instantaneous rate~$\delta_i$. Also, if {node~$i$ is susceptible and has an infected neighbor}, then the neighbor can infect node~$i$ with an instantaneous rate~$\beta_i$. The rates $\delta_i$ and $\beta_i$ are called the recovery and transmission rate of node $i$, respectively. {We suppose that the above model evolves in the continuous-time; therefore, the rates represent probabilities \emph{per unit
time}~\cite{Pastor-Satorras2015a} and, therefore, can be greater than one.}

\begin{figure}
\centering 
\includegraphics[width=.65\linewidth]{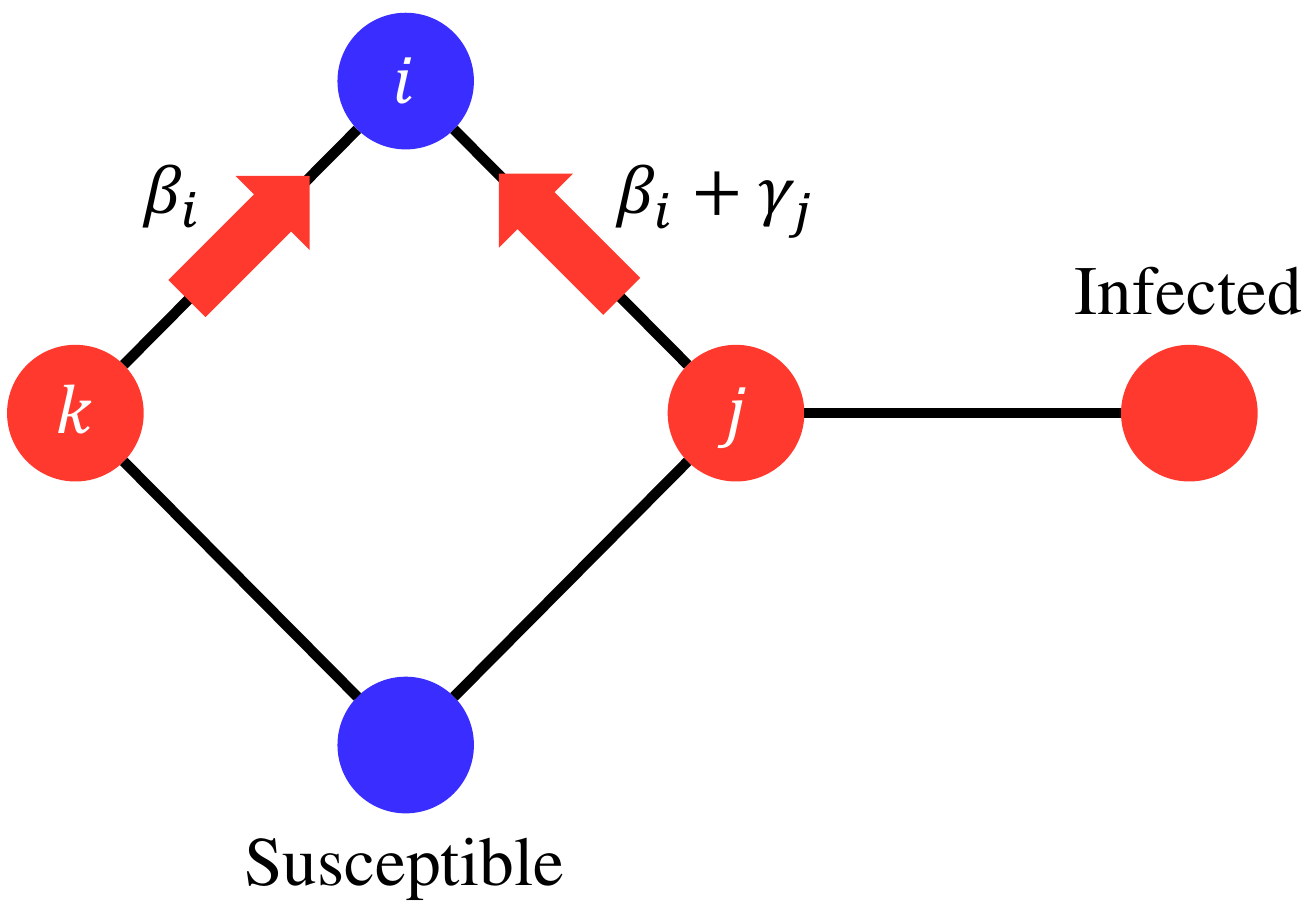}
\caption{Synergistic SIS model. Node $j$ infects node $i$ with an instantaneous rate $\beta_i + \gamma_j$ because node $j$ has one infected neighbor, while node $k$ infects node $i$ with an instantaneous rate $\beta_i$ as in the SIS model.}
\label{fig:SSIS}
\end{figure}

The SIS model uses a pair-wise description for infection events and, therefore, does not allow synergistic effects. In order to study the effect of synergy, we study the \emph{synergistic SIS model} \cite{Ludlam2012}, which we describe below. In the model, while an infected node recovers with the rate~$\delta_i$ as in the SIS model, the rate at which a susceptible node~$i$ is infected by its infected neighbor, say, node~$j$, is equal to
\begin{equation}\label{eq:synergyInfectionRate}
\beta_i + \gamma_j m_j(t), 
\end{equation} 
where $\gamma_j$ denotes the strength of synergy and is dependent on individual nodes, and $m_j(t)$ denotes the number of the infected neighbors of node~$j$ at time~$t$. The second term in~\eqref{eq:synergyInfectionRate} is introduced to describe synergistic effects: a susceptible node is directly affected not only by its infected neighbors but also their neighbors (see Fig.~\ref{fig:SSIS} for a schematic picture). We remark that, according to the classification in~\cite{Perez-Reche2011}, the synergistic SIS model incorporates only the \emph{d-synergies}, which are caused by adjacent pairs of infected nodes. Due to the limitations of the space, this brief does not deal with the other category of synergies called \emph{r-synergies} arising in the interactions between a susceptible node and its infected neighbors.

In this brief, we are interested in quantifying the rate at which the spreading process described by the synergistic SIS model grows. Therefore, we introduce the following definition:

\begin{definition}\label{defn:}
For all node~$i$ and time~$t\geq 0$, let $p_i(t)$ denote the probability that node $i$ is infected at time $t$. Define the \emph{growth rate} of the synergistic SIS model by $\rho = \sup_{V_0\subset \{1, \dotsc, n\}}\limsup_{t\to\infty}t^{-1}{\log \sum_{i=1}^N p_i(t)}$, where $V_0$ denotes the set of initially infected nodes.
\end{definition}

{In Definition~\ref{defn:}, the sum~$\sum_{i=1}^N p_i(t)$ represents the expected number of infected nodes. Therefore, the expression $\limsup_{t\to\infty}t^{-1}\log \sum_{i=1}^N p_i(t)$ measures the asymptotic growth of the expected number of infected nodes. Hence, the growth rate in Definition~\ref{defn:} quantifies the worst-case growth of the average size of the infected population.}

\begin{figure*}[!t] 
\normalsize
\setcounter{MYtempeqncnt}{\value{equation}}
\setcounter{equation}{8}
\begin{align}\label{eq:longDEp}
\frac{dp_i}{dt}&\leq -\delta_i p_i + \beta_i \sum_{j=1}^N a_{ij}p_j + \sum_{j=1 }^N \sum_{\substack{k<j\\k\neq i}} a_{ij}a_{kj}\gamma_{j} p_{kj} + \sum_{j=1 }^N \sum_{\substack{k>j\\k\neq i}} a_{ij}a_{kj}\gamma_{j} p_{jk},\\ 
\notag 
\frac{dp_{i\ell}}{dt}&\leq 
-\delta_\ell p_{i\ell} + \beta_{i\ell}(p_i-p_{i\ell}) + \sum_{m<i}a_{m\ell}\beta_\ell p_{mi} + \sum_{m > i}a_{m\ell}\beta_\ell p_{im} + \sum_{m=1}^{N}\Biggl(\sum_{\substack{
n<m\\n\neq \ell}}a_{m\ell}a_{nm}\gamma_{m}p_{nm} + \sum_{\substack{n>m\\n\neq \ell}}a_{m\ell}a_{nm}\gamma_{m}p_{mn}\Biggr) \\
&\quad \quad \quad - \delta_i p_{i\ell}+\beta_i a_{i\ell}(p_\ell - p_{i\ell}) + \sum_{j< \ell}a_{ji}\beta_i p_{j\ell} +  \sum_{j > \ell}a_{ji}\beta_i p_{\ell j}
+\sum_{j=1}^N \Biggl(\sum_{\substack{k<j\\ k\neq i}}a_{ji}a_{kj}\gamma_{j}p_{kj} + \sum_{\substack{k>j\\k\neq i}}a_{ji}a_{kj}\gamma_{j}p_{jk}\Biggr).\label{eq:longDEq}
\end{align}
\setcounter{equation}{\value{MYtempeqncnt}}
\hrulefill
\end{figure*}

The growth rate is closely related to fundamental notions in the network epidemiology such as  mean-time-to-absorption~\cite{Ganesh2005} and epidemic thresholds~\cite{VanMieghem2009a}. However, it is hard in practice to exactly compute the growth rate. Since the synergistic SIS model, as a Markov process, has a total of $2^N$ states, the computation of the growth rate requires finding the eigenvalues of a $2^N\times 2^N$ matrix representing the infinitesimal generator of the Markov process~\cite{VanMieghem2009a}. Therefore, the computational complexity of exactly calculating the growth rate increases exponentially with the number of the nodes \cite{Pan1999}. To overcome this difficulty, in the next section we introduce an upper-bound on the growth rate in terms of the eigenvalues of a matrix whose size grows only quadratically with $N$.

\section{Growth rate}\label{sec:grwoth}

In this section, we derive an upper bound on the growth rate of the synergistic SIS model. For each node~$i$ and time~$t$, let us define the variable $x_i(t)$ as $x_i(t)=1$ if node~$i$ is infected at time~$t$, and $x_i(t)=0$ otherwise. Then, since $m_j(t) = \sum_{k\in\mathcal N_j\backslash\{i\}}x_k(t)$, the synergistic rate of infection~\eqref{eq:synergyInfectionRate} shows that the transition probability of the variable $x_i$ is described as $\Pr(x_i(t+h) = 1 \mid x_i(t) = 0) = \sum_{j \in \mathcal N_i}x_j(t) ( \beta_i  + \gamma_j \sum_{k \in \mathcal N_j \backslash \{i\}} x_k(t)) h + o(h)$ and $\Pr(x_i(t+h) = 0 \mid x_i(t) = 1) = \delta_i h + o(h)$ for $h>0$. This implies that the variable~$x_i$ obeys the following stochastic differential equation with Poisson jumps (see, e.g., \cite{Hanson2007}):
\begin{equation}\label{eq:dxi=}
\begin{multlined}[.8\linewidth]
dx_i = -x_i d\Pi_{\delta_i} + (1-x_i)\sum_{j\in \mathcal N_i}
x_j d\Pi_{\beta_i} \\
+ (1-x_i) \sum_{j\in \mathcal N_i}\sum_{k\in \mathcal N_j\backslash\{i\}}x_jx_kd\Pi_{\gamma_{j}}, 
\end{multlined}
\end{equation}
where $\Pi_{\delta_i}$, $\Pi_{\beta_i}$, and $\Pi_{\gamma_{j}}$ are independent Poisson counters with rates~$\delta_i$, $\beta_i$, and~$\gamma_j$, respectively. In the stochastic differential equation~\eqref{eq:dxi=}, the first, second, and third terms represent the recovery, direct infection from neighbors, and synergy effects.

By \eqref{eq:dxi=}, the expectation $\mathbb E[x_i]$ satisfies
\begin{equation}\label{eq:dexi/dt}
\begin{multlined}[.8\linewidth]
\frac{d}{dt}\mathbb E[x_i] =
-\delta_i \mathbb E[x_i] + \sum_{j\in \mathcal N_i}\beta_i \mathbb E[(1-x_i)x_j]  \\+\sum_{j\in \mathcal N_i}\sum_{k\in \mathcal N_j\backslash\{i\}} \gamma_{j}\mathbb E[(1-x_i)x_jx_k]. 
\end{multlined}
\end{equation}
{Notice that we have $p_i(t) = \mathbb E[x_i(t)]$. Also, let} us introduce the notations $p_{ij}(t) = \mathbb E[x_i(t)x_j(t)]$ and $p_{ijk}(t) = \mathbb E[x_i(t)x_j(t)x_k(t)]$ for all $i, j, k \in \{1, \dotsc, N\}$ and $t\geq 0$. Then, the differential equation~\eqref{eq:dexi/dt} is rewritten as
\begin{equation}\label{eq:dpi/dt=}
\begin{multlined}[.8\linewidth]
\frac{dp_i}{dt} =
-\delta_i p_i +\sum_{j\in \mathcal N_i}\beta_i (p_j-p_{ij}) \\+ \sum_{j\in \mathcal N_i}\sum_{k\in \mathcal N_j\backslash\{i\}} \gamma_{j}(p_{jk}-p_{ijk}). 
\end{multlined}
\end{equation}
This differential equation rigorously describes the evolution of the infection probabilities~$p_i$, which we require to compute the growth rate~$\rho$. However, the differential equation contains the higher-order terms $p_{ij}$ and $p_{ijk}$ and, therefore, is not closed in the first-order term~$p_i$. For this reason, we cannot readily use the differential equation for calculating the growth rate.

We avoid this difficulty by ignoring the third order terms~$p_{ijk}$ and deriving a set of differential inequalities for upper-bounding the growth rate of the synergistic SIS model. Intuitively, we can expect that ignoring the higher-order terms brings relatively small errors in the regime of extinction (i.e., when the epidemics dies out), in which lower-order terms become dominant. From differential equation~\eqref{eq:dpi/dt=}, we obtain 
\begin{equation}\label{eq:dpi/dt}
\frac{dp_i}{dt}\leq -\delta_i p_i + \beta_i \sum_{j=1}^N a_{ij}p_j + \sum_{j=1 }^N \sum_{k\neq i} a_{ij}a_{kj}\gamma_{j} p_{jk} .
\end{equation}
It should be emphasized that, in deriving inequality~\eqref{eq:dpi/dt}, we chose to ignore the negative second order terms $-\beta_ip_{ij}$ for $j\in\mathcal N_i$ as well as the third-order terms. We discuss the reason for and consequence from this manipulation later in Remark~\ref{rem:}. 

Then, let us derive differential equations for the evolution of the second-order terms. If we apply It\^o's formula~\cite{Hanson2007} to the mapping $(x_i, x_\ell) \mapsto x_ix_\ell$ and, then, take mathematical expectations in the resulting stochastic differential equation, we obtain
\begin{equation}\label{eq:dexixell/dt}
\begin{multlined}[.85\linewidth]
\frac{d}{dt}\mathbb E[x_ix_\ell]=\!\!-\delta_\ell \mathbb E[x_ix_\ell]+\sum_{m=1}^N a_{m\ell}\beta_\ell \mathbb E[x_i(1-x_\ell)x_m] \\+ \sum_{m=1}^N \sum_{n\neq \ell}a_{m\ell}a_{nm} \gamma_{m}\mathbb E[x_i(1-x_\ell)x_m x_n] \\- \delta_i \mathbb E[x_i x_\ell] + \sum_{j=1}^Na_{ji}\beta_i \mathbb E[(1-x_i)x_jx_\ell] \\+ \sum_{j=1}^N\sum_{k\neq i}a_{ji}a_{kj} \gamma_{j} \mathbb E[(1-x_i)x_jx_kx_\ell]. 
\end{multlined}
\end{equation}
Since $x_i$ is $\{0, 1\}$-valued, the second term on the right-hand side of the differential equation~\eqref{eq:dexixell/dt} is bounded from above as $\sum_{m=1}^N a_{m\ell}\beta_\ell \mathbb E[x_i(1-x_\ell)x_m] \leq a_{i\ell}\beta_\ell (p_i-p_{i\ell}) + \sum_{m\neq i} a_{m\ell}\beta_\ell p_{im}$. Likewise, the fifth term on the right-hand side of \eqref{eq:dexixell/dt} can be bounded as $\sum_{j=1}^Na_{ji}\beta_i \mathbb E[(1-x_i)x_jx_\ell] \leq a_{\ell i}\beta_i (p_\ell-p_{i\ell})+\sum_{j\neq \ell}a_{ji}\beta_i p_{j\ell}$. Furthermore, we have
\begin{equation}\label{eq:ijkl}
\mathbb E[x_i(1-x_\ell)x_m x_n] \leq p_{mn},\ \mathbb E[(1-x_i)x_jx_kx_\ell]\leq p_{jk}. 
\end{equation}
These observations yield that 
\begin{equation}\label{eq:dpil/dt}
\begin{multlined}[.85\linewidth]
\frac{dp_{i\ell}}{dt}\leq 
-\delta_\ell p_{i\ell} + a_{i\ell}\beta_\ell(p_i-p_{i\ell}) \\+ \sum_{m\neq i}a_{m\ell}\beta_\ell p_{im} + \sum_{m=1}^{N}\sum_{n\neq \ell}a_{m\ell}a_{nm}\gamma_{m}p_{mn} \\- \delta_i p_{i\ell}+ a_{\ell i}\beta_i(p_\ell - p_{i\ell}) \\+ \sum_{j\neq \ell}a_{ji}\beta_i p_{j\ell}
+\sum_{j=1}^N \sum_{k\neq i}a_{ji}a_{kj}\gamma_{j}p_{kj}.
\end{multlined}
\vspace{3mm}
%\end{aligned}
\end{equation}

\begin{remark}\label{rem:alternativebounds}
One can derive different bounds on the derivative $dp_{i\ell}/dt$ by replacing the reductions in~\eqref{eq:ijkl} with others such as $\mathbb E[x_i(1-x_\ell)x_m x_n] \leq p_{im}$ and $\mathbb E[(1-x_i)x_jx_kx_\ell]\leq p_{j\ell}$. It is not theoretically easy to identify the best reduction giving the tightest bound on the derivative. However, we have numerically observed that using different reductions does not significantly alter the results of numerical simulations performed in Section~\ref{sec:sim}. 
\end{remark}

\newcommand{\partialubarESIS}{$\partial\mbox{\underbar{$\mathcal E$}}_{\SIS}$}
\newcommand{\ubarE}{\underbar{${\mathcal E}$} }
\newcommand{\partialubarE}{$\partial$\underbar{${\mathcal E}$} }

\begin{figure*}%
\centering
\includegraphics[width=1\linewidth]{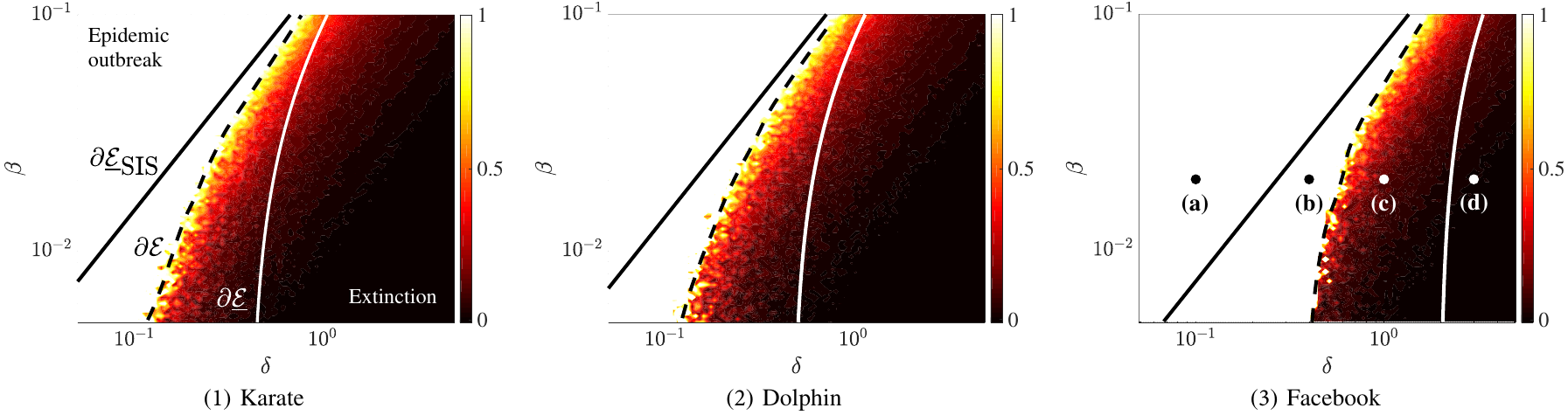}
\caption{Meta-stable number $y^\star$ of infected nodes in the synergistic SIS model for various values of $\delta$ and $\beta$. Colored region: the numerically obtained extinction region~$\mathcal E$. Dashed black line: the boundary $\partial \mathcal E$ of $\mathcal E$. White (black) solid line: the boundary~\partialubarE (\partialubarESIS, respectively) of the extinction region estimated from Theorem~\ref{thm:} (by the conventional method, respectively). Since the boundary \partialubarESIS lies in the region of epidemic outbreak, the conventional method overestimates the extinction region, while our estimate~\ubarE is contained in $\mathcal E$ for all networks.} 
\label{3figs}
\vspace{6mm}
\includegraphics[width=1\linewidth]{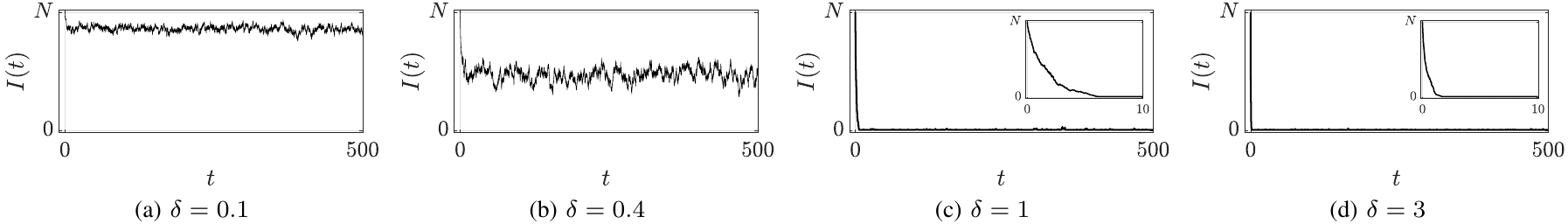}
\caption{A sample path of the number of infected nodes, denoted by $I(t)$, versus time~$t$ in the Facebook network. The transmission rate is fixed to be $\beta = 0.02$, while the recovery rates are changed as $\delta=0.1$, $0.4$, $1$, and $3$. The corresponding points $(\delta, \beta)$ are plotted in Fig.~2.3.} 
\label{fig:4figs}
\end{figure*}

By replacing second-order moments~$p_{ij}$ having indices~$i>j$ with their equivalent counterparts~$p_{ji} = E[x_jx_j] = E[x_ix_j] = p_{ij}$, we rewrite the differential inequality~\eqref{eq:dpi/dt} as \eqref{eq:longDEp} for all $i$. In the same manner, we rewrite the differential inequality \eqref{eq:dpil/dt} as \eqref{eq:longDEq}, which contains only the variables belonging to the  set $\mathcal P = \{ p_i, p_{jk}\colon  i, j, k\in \{1, \dotsc, N\} \mbox{ and }j < k \}$. Now, let $p$ denote the $\frac{N^2+N}{2}$-dimensional vectorial variable that can be obtained by stacking the variables in the set~$\mathcal P$. Then, we rewrite the system of differential inequalities~\eqref{eq:longDEp} and~\eqref{eq:longDEq}\setcounter{equation}{10} as 
\begin{equation}\label{eq:dpdtleq}
\frac{dp}{dt} \leq \mathcal M p
\end{equation}
for a uniquely determined $\frac{N^2+N}{2}\times \frac{N^2+N}{2}$ matrix  $\mathcal M$. 

\begin{remark}\label{rem:}
Since we ignored second order terms~$-\beta_ip_{ij}$ for deriving~\eqref{eq:dpi/dt}, one should expect a certain conservativeness in the bound~\eqref{eq:dpdtleq} and, therefore, the bound of the decay rate to be later derived in Theorem~\ref{thm:}. However, it is necessary to ignore the negative second order terms because, if not ignored, we cannot use the fact that the matrix~$\mathcal M$ is Metzler for the proof of the theorem. 
\end{remark}

Inequality~\eqref{eq:dpdtleq} shows that, for all $t\geq 0$, there exists a nonnegative vector $\epsilon(t)$ such that ${dp}/{dt} = \mathcal M p - \epsilon$. From differential inequalities~\eqref{eq:longDEp} and~\eqref{eq:longDEq}, we see that the off-diagonals of $\mathcal M$ are nonnegative, i.e., $\mathcal M$ is a Metzler matrix. This implies that the exponential~$\exp(\mathcal M t)$ has only nonnegative entries for all $t\geq 0$ (see, e.g.,~\cite{Ogura2015i}). Furthermore, by definition, each entry of the vector~$\epsilon(t)$ is nonnegative for all $t\geq 0$. Therefore, we obtain $p(t) = \exp(\mathcal M t)p(0) - \int_0^t \exp(\mathcal M(t-\tau))\epsilon(\tau)\,d\tau \leq  \exp(\mathcal M t)p(0)$. This inequality bounds the vector~$p$, which contains the infection probabilities $p_1$, \dots,~$p_N$, by the matrix exponential~$\exp(\mathcal M t)$. We have thus proved the following theorem, which is the main theoretical result of this brief:

\begin{theorem}\label{thm:}
The growth rate $\rho$ of the synergistic SIS model satisfies $\rho \leq \lambda_{\max}(\mathcal M)$. 
\end{theorem}

\section{Numerical simulations}\label{sec:sim}

In this section, we illustrate the effectiveness of Theorem~\ref{thm:} by numerical simulations. From Theorem~\ref{thm:}, we can obtain a sufficient condition for the spreading process to become extinct (i.e., the spread is not invasive) as 
\begin{equation}\label{eq:sufficient}
\lambda_{\max}(\mathcal M)< 0. 
\end{equation}
This section is devoted to numerically validating the sufficiency of the extinction condition~\eqref{eq:sufficient}. To determine if the spreading process becomes extinct, we use the notion of the \emph{meta-stable} number of infected nodes that is commonly used in the Network Science. To identify the meta-stable number of infected nodes, we begin by computing the following long-time average of the number of infected nodes $y(t) = t^{-1}\int_0^t \sum_{i=1}^N x_i(\tau) \,d\tau$ for a sufficiently large $t$. In simulations, the spreading process can die out due to random fluctuations. To prevent this from happening, we use the re-infection procedure~\cite{Cator2013a} in which a randomly chosen node is infected immediately after the spreading process dies. After we finish the simulation, we determine the meta-stable number of infected nodes, denoted by~$y^\star$, by
\begin{equation*}
y^* = y(t) - 1,
\end{equation*}
where we subtract one to compensate for the effect of the re-infection procedure. Finally, when $y^\star < 1$, it is determined that the spreading process becomes extinct.

In our numerical simulations, we use the following three empirical human and animal social networks: (1) a social network of Zachary's Karate club~\cite{Zachary1977} (\emph{Karate}, $N = 34$), (2) a social network of bottlenose dolphins~\cite{Kunegis2013} (\emph{Dolphin}, $N = 62$), and (3) a friendship network in Facebook (\emph{Facebook}, $N=247$). For simplicity of illustration, we assume that the network consists of homogeneous nodes; i.e., we assume $\beta_i = \beta$, $\delta_i = \delta$, and $\gamma_{j} = \gamma$ for constants $\beta$, $\delta$, and $\gamma$. In this numerical simulation, we fix $\gamma = 10^{-2}$. For the above three networks and various values of the transmission rate~$\beta$ and the recovery rate~$\delta$, we compute the meta-stable number $y^\star$ of infected nodes using the procedure described above, as well as our upper-bound~$\lambda_{\max}(\mathcal M)$ on the growth rate. We then obtain the true extinction region
\begin{equation*}
\mathcal E = \{(\delta, \beta)\colon y^\star < 1\}
\end{equation*}
and its inner estimate from the sufficient condition~\eqref{eq:sufficient}:
\begin{equation*}
\ubar{\mathcal E} = \{(\delta, \beta)\colon \lambda_{\max}(\mathcal M) < 0\}.
\end{equation*}

In Fig.~\ref{3figs}, we illustrate the regions $\mathcal E$ and $\ubar{\mathcal E}$ for the three empirical networks. We confirm that, for any of the networks, our inner-estimate~$\ubar{\mathcal E}$ is contained in the true extinction region~$\mathcal E$ (i.e., the boundary~$\partial \ubar{\mathcal E}$ indicated by the white line passes through the colored region). For comparison, in Fig.~\ref{3figs}, we also illustrate the estimate
\begin{equation*}
\ubar{\mathcal E}_{\SIS} = \{(\delta, \beta)\colon\bar \rho_{\SIS} < 0\}
\end{equation*}
of the extinction region based on the conventional upper-bound $\bar \rho_{\SIS} =  \beta \lambda_{\max}(A) -\delta$ on the growth rate of the SIS model without synergy \cite{VanMieghem2009a}. Since the boundary~$\partial \ubar{\mathcal E}_{\SIS}$  passes outside the true extinction region for any of the three graphs, we observe that the use of the conventional method, which does not take into account synergistic effects, can lead to the overestimate of the extinction regions. In Fig.~\ref{fig:4figs}, we show  sample paths of the time-evolution of the number of infected nodes for $\beta=0.02$ and $\delta=0.1$, $0.3$, $1$, and $3$. From Fig.~\ref{fig:4figs}.b, we confirm that the conventional method does overestimate the region of extinction.

\section{Conclusion}

In this brief, we have studied the synergistic SIS model over arbitrary complex networks. Since the state space of the model grows exponentially fast with the number of nodes, it is almost impossible to directly analyze the model except when the network size is very small. To overcome this difficulty, we have derived an upper-bound on the growth rate the spreading model in terms of the eigenvalues of a matrix whose size grows quadratically with the number of the nodes. We have illustrated the effectiveness of our theoretical results by numerical simulations on empirical human and animal social networks. A possible direction for future research is eliminating the computational complexity for calculating our upper-bound. Since the size of $\mathcal M$ grows quadratically, it is still not easy to apply the current bound for large-scale networks. It is of theoretical interest to examine if we can exploit the sparsity structure of $\mathcal M$ to efficiently calculate the upper bound.


\end{document}